\documentclass[onecolumn,tightenlines,nofootinbib,preprintnumbers,
amsmath,amssymb]{revtex4}

\usepackage[usenames,dvipsnames]{color}
\usepackage{caption2}
\usepackage{dcolumn}
\usepackage{graphicx}
\usepackage{subfigure}
\usepackage{epstopdf}
\usepackage{bm}
\usepackage{booktabs}
\usepackage{lscape}
\usepackage{hyperref}
\usepackage{cleveref}
\renewcommand{\arraystretch}{0.9}
\usepackage{setspace}

\begin{document}
\begin{spacing}{1.5}

\title{
CP violation induced by the real part of the interference term in $\rho^0 - \omega$ mixing  }

\author{Jin-Zhao Guo $^{1}$, Gang L\"{u}$^{1}$\footnote{Email: ganglv66@sina.com}, Zhen-Hua Zhang$^{2}$}

\affiliation{\small $^{1}$School of Physics and Advanced Energy, Henan University of Technology, Zhengzhou 450001, China\\
	\small $^{2}$ School of Nuclear Science and Technology, University of South China, Hengyang 421001, China
}

\begin{abstract}
To circumvent the severe numerical cancellations in the standard integrated CP asymmetry ($A_{CP}$) near the $\omega$ mass $m_\omega$, we propose a modified CP-violating observable, $A_{CP}^{\Re}$, which explicitly highlights the contribution of the real part of the interference term. Subsequently, we employ the three-body hadronic decay $B^{-} \rightarrow \pi^{+} \pi^{-} \pi^{-}$ within the Perturbative QCD (PQCD) approach as a primary case study to validate this theoretical framework. Furthermore, this method naturally eliminates the smooth, sign-preserving continuum background originating from broad scalar resonances like the $f_0(500)$. This generalized framework provides clean and robust theoretical guidance for recovering localized CP-violation signals that might otherwise be masked by coarse experimental binning at future high-luminosity colliders.
\end{abstract}

\maketitle

\section{Introduction}
\label{sec:sample1}

CP violation (CPV) plays a crucial role in both testing the Standard Model (SM) of particle physics and in probing potential new physics. Experimentally, CPV has been observed in the decay processes of \textit{K}, \textit{D}, and \textit{B} mesons \cite{KTeV:1999aiu,BaBar:2001oxa,Belle:2001zzw,LHCb:2019hjh}. The LHCb experiment has reported significant CP violation in localized regions of the decays $B^{\pm} \rightarrow K^{\pm} \pi^{+} \pi^{-}$ and $B^{\pm} \rightarrow K^{\pm} K^{+} K^{-}$ \cite{LHCb:2013ptu}. Moreover, experimental studies have identified multiple sources of CP violation in the decay $B^{-} \rightarrow \pi^{+} \pi^{-} \pi^{-}$ \cite{LHCb:2019jta,LHCb:2019sus}. Particularly noteworthy is the recent observation by the LHCb Collaboration of CP violation in the multibody decay $\Lambda_{b}^{0} \rightarrow pK^{-}\pi^{+}\pi^{-}$, which marks the first observation of CP violation in the baryon sector \cite{LHCB-bary2025}. These discoveries provide important direction and scientific basis for in-depth research on the theoretical mechanism of CP violation.

The theoretical description of CPV in multibody decays faces significant challenges, mainly due to the combined influence of the weak phase difference from the CKM matrix and the strong phase difference between tree diagrams and penguin diagrams \cite{Yan:2023yvx}. The contributions from strong phases involve hadronic matrix elements, interference effects among intermediate resonances, and final-state interactions. Whether the decay process involves mesons or baryons, the complex interference effects resulting from multiple intermediate resonances in multibody decays make the prediction of CPV particularly difficult \cite{Yan:2022kck}.

 In the decay networks of heavy hadrons—encompassing $B$ mesons, $D$ mesons, and heavy baryons—multiple intermediate states are routinely produced, generating intricate interference structures. These intermediate states will produce complex interference effects in the resonance region. Based on the vector meson dominance model (VMD), 
the photon propagator couples to vector mesons \cite{VMD1969}. Therefore, $\rho^0 - \omega$  mixing is proposed and applied to the study of direct CP violation in hadronic B decays \cite{Gardner1998}.
The CP violation mechanism mediated by $\rho^0 - \omega$ mixing in hadronic B decays has been investigated in recent years, with studies revealing a significant  CP violation within the interference region \cite{Guo2000, yuting2018, Li2019}.

Recently, a mechanism for generating CP violation was proposed, based on the imaginary part arising from the interference of two amplitudes \cite{zhangzhenhua2024,Durieux:2015zwa}. The interference between $\rho^0$ and $\omega$ mesons provides an excellent platform to test this mechanism, due to the nearly degenerate masses of the $\rho^0$ and $\omega$ mesons, thereby enabling more accurate theoretical predictions and offering valuable guidance for experimental investigations.

The structure of this paper is organized as follows: In Sec. \ref{Sec 2}, we present a detailed computational formulation of the physical mechanism involved in the $\rho^0 - \omega$ mixing. In Sec. \ref{Sec 3}, we analyze the contribution of interference effects arising from this mechanism in the decay process $B^{-} \rightarrow \pi^{+}\pi^{-}\pi^{-}$ within the framework of Perturbative QCD (PQCD).  In Sec. \ref{Sec 4}, we present relevant theoretical parameters and numerical results for CP violation. Finally, a summary and discussion are given in Sec. \ref{Sec 6}.

\section{Detailed Computational Formulation of the Physical Mechanism Involved in the $\rho^0 - \omega$ Mixing}\label{Sec 2}

We consider the interference of two intermediate resonant states, namely the $\rho^0$ and $\omega$ mesons, within specific and localized regions of the phase space. To the first order in isospin violation \cite{Guo:2000uc}, when the invariant mass $\sqrt{s}$ is close to the $\omega$ resonance mass, unlike the simple two-particle interference \cite{zhangzhenhua2024}, the modulus square of the interference amplitude via $\rho^0 - \omega$ mixing can be expressed as:
\begin{equation}
	\begin{split}
		\begin{aligned}
			\overline{|\mathcal{M} |^{2}}=\frac{1}{\left|s_{\rho^{0}} \right|^{2}}\left(\left|\mathcal{A}\right|^{2}+2 \Re\left(\frac{\mathcal{A}^{*} \mathcal{B}\tilde{\Pi}_{\rho\omega}}{s_{\omega}}\right)+\frac{|\mathcal{B}\tilde{\Pi}_{\rho\omega}|^{2}}{\left|s_{\omega}\right|^{2}}\right),
		\end{aligned}
	\end{split}
	\label{Eq.2}
\end{equation}
where $\mathcal{A}$ and $\mathcal{B}$ represent the amplitudes for the cascade processes $H \rightarrow (\rho^0 \rightarrow \pi^+\pi^-) X$ and $H \rightarrow (\omega-\rho^0 \rightarrow \pi^+\pi^-) X$, respectively, without incorporating the corresponding propagators. Here, $H$ and $X$ denote the parent particle and the accompanying spectator particle, respectively, which are required to share the same charge sign. The amplitude $\mathcal{A}$ encompasses the decay process of the intermediate $\rho^0$ meson into two $\pi$ mesons. 
The direct decay $\omega \rightarrow \pi^{+}\pi^{-}$ is effectively incorporated into $\tilde{\Pi}_{\rho\omega}$ \cite{Lu:2010vb}, resulting in an explicit $s$-dependence of $\tilde{\Pi}_{\rho\omega}$, where  
$\widetilde{\Pi}_{\rho \omega} = \frac{s_{\rho} \Pi_{\rho \omega}}{s_{\rho} - s_{\omega}}$ \cite{Maltman1996,changchang}. The term $s_{r}$ (with $r = \rho$ or $\omega$) represents the reciprocal of the Breit-Wigner propagator, which can be explicitly written as $s_{r} = s - m_{r}^{2} + i m_{r} \Gamma_{r}$. $\mathcal{A}^{*}$ is the complex conjugate of ${\mathcal{A}}$. In this expression, $\sqrt{s}$ represents the invariant mass of the $\pi^{+}\pi^{-}$ system, whereas $m_r$ and $\Gamma_{r}$ correspond to the mass and decay width of the resonance $r$, respectively. When performing integration within the interference region, the condition $\left|s_{\rho}\right| \gg \left|s_{\omega}\right|$ is satisfied, which yields the approximation $\widetilde{\Pi}_{\rho \omega} \approx \Pi_{\rho \omega} = \left|\Pi_{\rho \omega}\right| e^{i \theta}$. In the subsequent sections, $\widetilde{\Pi}_{\rho \omega}$ is consistently replaced by $\Pi_{\rho \omega}$.
The overline placed over the squared decay amplitude indicates that a summation over the spin or helicity states of both the initial and final particles may be performed.
$\Re$ represents the value of the real part within the parentheses.

For the convenience of analysis, we represent the interference term from Eq. \eqref{Eq.2} as
\begin{equation}
	\begin{split}
		\begin{aligned}
			\Re\left(\frac{\mathcal{A}^{*} \mathcal{B}\Pi_{\rho \omega}}{s_{\omega}}\right)=\frac{\Im\left(\mathcal{A}^{*} \mathcal{B}\Pi_{\rho \omega}\right) \Im\left(s_{\omega}\right)+\Re\left(\mathcal{A}^{*} \mathcal{B}\Pi_{\rho \omega}\right) \Re\left(s_{\omega}\right)}{\left|s_{\omega}\right|^{2}},
		\end{aligned}
	\end{split}
	\label{Eq.3}
\end{equation}
 where  $\Im$ represents the imaginary part  within the parentheses. It can be seen that the interference term has been divided into two terms. These two terms are respectively proportional to $\Re\left(s_{\omega}\right)=s-m_{\omega}^{2}$ and $\Im\left(s_{\omega}\right)=m_{\omega} \Gamma_{\omega}$.
Furthermore, we are concerned with the interference effect, so the invariant mass $\sqrt{s}$ needs to be confined within the interference region. Since $\Gamma_{\rho}>\Gamma_{\omega} $, the interference region should be set as $({m_{\omega}{\pm}\Gamma_{\omega}})^2$.
The CP violation can be defined as:
\begin{equation}
	\begin{split}
		A_{C P}\equiv \frac{\int_{({m_{\omega}-\Delta_{-}})^2}^{({m_{\omega}+\Delta_{+}})^2}\left(\overline{|\mathcal{M}|^{2}}-\overline{|\overline{\mathcal{M}}|^{2}}\right)  d s}{\int_{({m_{\omega}-\Delta_{-}})^2}^{({m_{\omega}+\Delta_{+}})^2}\left(\overline{|\mathcal{M}|^{2}}
			+\overline{|\overline{\mathcal{M}}|^{2}}\right) d s},
	\end{split}
	\label{Eq.5}
\end{equation}
where $\overline{\mathcal{M}}$ denotes the decay amplitude of the CP-conjugate process, and $\Delta_{-} = \Delta_{+} = \Gamma_{\omega}$. At this point, the decay width of the $\rho^{0}$ meson encompasses that of the $\omega$ meson. Consequently, $\Delta_{-}$ and $\Delta_{+}$ are equal. However, in cases where two interfering particles do not fully overlap, $\Delta_{-}$ and $\Delta_{+}$ will no longer be equal.

By combining formulas \eqref{Eq.2}-\eqref{Eq.5}, the numerator of $A_{CP}$ can be decomposed into four components, namely $I_{1}^{\left(\Delta_{-}, \Delta_{+}\right)}$, $I_{2}^{\left(\Delta_{-}, \Delta_{+}\right)}$, $I_{3}^{\left(\Delta_{-}, \Delta_{+}\right)}$, and $I_{4}^{\left(\Delta_{-}, \Delta_{+}\right)}$. Dividing the numerator into these four parts is helpful for systematically analyzing the various factors influencing $A_{CP}$ :  
\begin{equation}  
	\begin{split}  
		A_{C P} \propto I_{1}^{\left(\Delta_{-}, \Delta_{+}\right)} + I_{2}^{\left(\Delta_{-}, \Delta_{+}\right)} + I_{3}^{\left(\Delta_{-}, \Delta_{+}\right)} + I_{4}^{\left(\Delta_{-}, \Delta_{+}\right)},
	\end{split}  
\end{equation}  
which are explicitly defined as follows:
\begin{equation}
	\begin{array}{l}
		I_{1}^{\left(\Delta_{-}, \Delta_{+}\right)} \equiv 	{\mathcal{N}_{\left|\mathcal{A}\right|^{2}}}\int_{({m_{\omega}-\Delta_{-}})^2}^{({m_{\omega}+\Delta_{+}})^2} d s,
	\end{array}
	\label{Eq.7}	
\end{equation}
\begin{equation}
	\begin{array}{l}
		I_{2}^{\left(\Delta_{-}, \Delta_{+}\right)} \equiv	{\mathcal{N}_{\left|\mathcal{B}\right|^{2}}} \int_{({m_{\omega}-\Delta_{-}})^2}^{({m_{\omega}+\Delta_{+}})^2} \frac{ \left|\Pi_{\rho \omega}\right|^{2}}{\left|s_{\omega}\right|^{2}} d s,
	\end{array}
\end{equation}
\begin{equation}
	\begin{array}{l}
		I_{3}^{\left(\Delta_{-}, \Delta_{+}\right)} \equiv  2\mathcal{N}_{\Im\left(\mathcal{A}^*\mathcal{B}e^{i\theta}\right)} \int_{({m_{\omega}-\Delta_{-}})^2}^{({m_{\omega}+\Delta_{+}})^2} \frac{m_{\omega} \Gamma_{\omega}\left|\Pi_{\rho \omega}\right|}{\left|s_{\omega}\right|^{2}} d s,\\
		
	\end{array}
\end{equation}
and
\begin{equation}
	\begin{array}{l}
		I_{4}^{\left(\Delta_{-}, \Delta_{+}\right)} \equiv 	2\mathcal{N}_{\Re\left(\mathcal{A}^*\mathcal{B}e^{i\theta}\right)}\int_{({m_{\omega}-\Delta_{-}})^2}^{({m_{\omega}+\Delta_{+}})^2}\frac{ \left(s-m_{\omega}^{2}\right)\left|\Pi_{\rho \omega}\right|}{\left|s_{\omega}\right|^{2}} d s,
	\end{array}
	\label{Eq.10}
\end{equation}
where $\mathcal{N}_\xi \equiv \xi - \overline{\xi}$ with $\xi \in \left\{ |\mathcal{A}|^2, \; |\mathcal{B}|^2, \; \Im(\mathcal{A}^*\mathcal{B}e^{i\theta}), \; \Re(\mathcal{A}^*\mathcal{B}e^{i\theta}) \right\}$.  $\overline{\mathcal{A}} \text { and } \overline{\mathcal{B}}$ are the corresponding amplitudes for the
CP-conjugate process.

It is evident that $I_{4}^{\left(\Delta_{-}, \Delta_{+}\right)}$ is a particularly significant component.   When $ s $ passes through $ m_{\omega}^{2} $, its sign changes, leading to the cancellation of the integral values of $ I_{4}^{(\Delta_{-}, \Delta_{+})} $ near $ m_{\omega}^{2} $. If the influence of $I_{4}^{\left(\Delta_{-}, \Delta_{+}\right)}$ within the $A_{C P}$ is substantial, the theoretically calculated CP violation value becomes significantly diminished due to this cancellation effect. 
 The analysis indicates that the $\rho^0-\omega$ mixing leads to the cancellation of the imaginary part of the cross term. To preserve the contribution of the real part of the interference term, we introduce a sign factor into the CP-violating observable:
\begin{equation}
	\begin{split}
		A_{C P}^{\Re} \equiv \frac{\int_{({m_{\omega}-\Delta_{-}})^2}^{({m_{\omega}+\Delta_{+}})^2}\left(\overline{|\mathcal{M}|^{2}}-\overline{|\overline{\mathcal{M}}|^{2}}\right) \operatorname{sgn}\left(s-m_{\omega}^{ 2}\right) d s}{\int_{({m_{\omega}-\Delta_{-}})^2}^{({m_{\omega}+\Delta_{+}})^2}\left(\overline{|\mathcal{M}|^{2}}+\overline{|\overline{\mathcal{M}}|^{2}}\right) d s},
	\end{split}
	\label{Eq.11}
\end{equation}
where the sign function $\operatorname{sgn}(x)$ is defined as
\begin{equation}
	\begin{split}
		\operatorname{sgn}(x)=\left\{\begin{array}{ll}
			+1, & x>0 \\
			-1, & x<0.
		\end{array}\right.
	\end{split}
\end{equation}
Therefore, the sign of $I_{4}^{\left(\Delta_{-}, \Delta_{+}\right)}$ remains unchanged before and after $\sqrt{s}$ passes through the $\omega$ mass ($m_\omega$), thus preventing potential cancellations during the integration process.

The interference effect in the $B^{-}  \rightarrow \pi^{+} \pi^{-} \pi^{-} \left(\operatorname{via} \: \rho^{0}-\omega\: \operatorname{mixing}\right)$ decay channel is primarily responsible for the sign flip of $A_{CP}$ within this kinematic region, which is attributed to the contribution of the $I_{4}^{\left(\Delta_{-}, \Delta_{+}\right)}$ term. Therefore, it is justifiable to incorporate the sgn sign function when the value of $I_{4}^{\left(\Delta_{-}, \Delta_{+}\right)}$ is relatively large. In order to facilitate the direct comparison of $I_{4}^{\left(\Delta_{-}, \Delta_{+}\right)}$ with $I_{1}^{\left(\Delta_{-}, \Delta_{+}\right)}$, $I_{2}^{\left(\Delta_{-}, \Delta_{+}\right)}$, and $I_{3}^{\left(\Delta_{-}, \Delta_{+}\right)}$, we define as follows:

\begin{equation}
	\begin{split}
		I_{4}^{\left(\Delta_{-}, \Delta_{+}\right){\Re} }\equiv 	2\mathcal{N}_{\Re\left(\mathcal{A}^*\mathcal{B}e^{i\theta}\right)}\int_{({m_{\omega}-\Delta_{-}})^2}^{({m_{\omega}+\Delta_{+}})^2}\frac{ \left(s-m_{\omega}^{2}\right)\left|\Pi_{\rho \omega}\right|\operatorname{sgn}\left(s-m_{\omega}^{2}\right)}{\left|s_{\omega}\right|^{2}} d s,
	\end{split}
\end{equation}
and the amplitudes of $\mathcal{A}$ and $\mathcal{B}$ can be written as
\begin{equation}
	\begin{split}	
		\begin{array}{l}
			\mathcal{A}=\left|T_{\rho}\right| e^{i\left(\phi_{T \rho}+\delta_{T \rho}\right)}+\left|P_{\rho}\right| e^{i\left(\phi_{P \rho}+\delta_{\mathrm{P} \rho}\right)}, \\
			{\mathcal{B}}=\left|T_{\omega}\right| e^{i\left(\phi_{T \omega}+\delta_{T \omega}\right)}+\left|P_{\omega}\right| e^{i\left(\phi_{P \omega}+\delta_{\mathrm{P} \omega}\right)}, \\
			\bar{	\mathcal{A}}=\left|T_{\rho}\right| e^{i\left(\delta_{T \rho}-\phi_{T \rho}\right)}+\left|P_{\rho}\right| e^{i\left(\delta_{\mathrm{P} \rho}-\phi_{P \rho}\right)}, \\
			\bar{\mathcal{B}}=\left|T_{\omega}\right| e^{i\left(\delta_{T \omega}-\phi_{T \omega}\right)}+\left|P_{\omega}\right| e^{i\left(\delta_{\mathrm{P} \omega}-\phi_{P \omega}\right)},
		\end{array}	
	\end{split}
	\label{Eq.15}	
\end{equation}
 where $T_{\mathrm{i}}$ represents the contribution from the tree-level diagram, $\mathrm{P}_{\mathrm{i}}$ denotes the contribution from the penguin diagram, $\phi$ signifies the weak phase, and $\delta$ indicates the strong phase. 
For the two decay channels of $B^{-} \rightarrow \left(\rho^0 \rightarrow \pi^{+}\pi^{-}\right)\pi^{-}$ and $B^{-} \rightarrow \left(\omega -\rho^{0} \rightarrow \pi^{+} \pi^{-} \right) \pi^{-}$, their weak phases are identical, and thus the conditions $\phi_{T \rho}=\phi_{T \omega}$ and $\phi_{P \rho}=\phi_{P \omega}$ are satisfied.  Based on this information, we can proceed to calculate the values of $\mathcal{N}_{\left|\mathcal{A}\right|^{2}}$, $\mathcal{N}_{\left|\mathcal{B}\right|^{2}} $, $\mathcal{N}_{\Im\left(\mathcal{A}^*\mathcal{B}e^{i\theta}\right)}$, and $\mathcal{N}_{\Re\left(\mathcal{A}^*\mathcal{B}e^{i\theta}\right)}$. 
\begin{equation}
	\begin{split}
		\begin{array}{l}
			\mathcal{N}_{\left|\mathcal{A}\right|^{2}} \equiv \left|T_{\rho}P_{\rho}\right|\left[-4 \sin \left(\delta_{T \rho}-\delta_{\mathrm{P} \rho}\right) \sin \left(\phi_{T \rho}-\phi_{P \rho}\right)\right] ,
		\end{array}
	\end{split}
\end{equation}
\begin{equation}
	\begin{split}
		\begin{array}{l}
	\mathcal{N}_{\left|\mathcal{B}\right|^{2}} \equiv  \left|T_{\omega}P_{\omega}\right|\left[-4 \sin \left(\delta_{T \omega}-\delta_{\mathrm{P} \omega}\right) \sin \left(\phi_{T \omega}-\phi_{P \omega}\right)\right] ,
		\end{array}
	\end{split}
\end{equation}
\begin{equation}
	\begin{aligned}
	\mathcal{N}_{\Im\left(\mathcal{A}^*\mathcal{B}e^{i\theta}\right)}\equiv  2 \sin \left(\phi_{T \rho}-\phi_{P \rho}\right)\left[-\left|T_{\rho} P_{\omega}\right| 
		\cos \left(\delta_{T \rho}-\delta_{\mathrm{P} \omega}-\theta\right)
		+\left|P_{\rho} T_{\omega}\right| \cos \left(\delta_{\mathrm{P} \rho}-\delta_{T \omega}-\theta\right)\right]	,
	\end{aligned}
\end{equation}
and
\begin{equation}
	\begin{aligned}
	\mathcal{N}_{\Re\left(\mathcal{A}^*\mathcal{B}e^{i\theta}\right)}\equiv  2\sin \left(\phi_{T \rho}-\phi_{P\rho}\right)\left[-\left|T_{\rho} P_{\omega}\right| \sin \left(\delta_{T \rho}-\delta_{\mathrm{P} \omega}-\theta\right)+\left|P_{\rho} T_{\omega}\right| \sin \left(\delta_{\mathrm{P} \rho}-\delta_{T \omega}-\theta\right)\right]  .
	\end{aligned}
\end{equation}
To isolate the components associated with amplitude contribution and phase, we introduce the following definition: 
\begin{equation}
	\begin{split}
		\int_{({m_{\omega}-\Delta_{-}})^2}^{({m_{\omega}+\Delta_{+}})^2} d s=J_1 ,\quad & \quad\quad\quad \int_{({m_{\omega}-\Delta_{-}})^2}^{({m_{\omega}+\Delta_{+}})^2}\frac{\left|\Pi_{\rho \omega}\right|^{2}}{\left|s_{\omega}\right|^{2}} d s=J_2, \\
		\\
		\int_{({m_{\omega}-\Delta_{-}})^2}^{({m_{\omega}+\Delta_{+}})^2} \frac{m_{\omega} \Gamma_{\omega}\left|\Pi_{\rho \omega}\right|}{\left|s_{\omega}\right|^{2}} d s=J_3, \quad&\quad\int_{({m_{\omega}-\Delta_{-}})^2}^{({m_{\omega}+\Delta_{+}})^2} \frac{\left(s-m_{\omega}^{2}\right)\left|\Pi_{\rho \omega}\right|\operatorname{sgn}\left(s-m_{\omega}^{2}\right)}{\left|s_{\omega}\right|^{2}} d s=J_4.
	\end{split}
\end{equation}

Here, $J_i$ ($i=1, 2, 3, 4$) represent the characteristic parameters governing the $\rho^0-\omega$ mixing system decaying into $\pi^+\pi^-$. Different choices of the parent hadrons $H$ and the accompanying bachelor mesons $X$ only alter the amplitude magnitudes as well as the strong and weak phases, thereby leading to different numerical values for $\mathcal{N}_\xi$, while leaving the values of $J_i$ entirely unaffected. Evidently, the numerical values of $J_i$ depend strictly on the choice of the integration region; in this work, the integration window is specified as $m_{\omega} \pm \Gamma_{\omega}$. Consequently, these $J_i$ parameters can be directly evaluated using numerical computation software. Our numerical calculation yields $J_1=0.0266$, $J_2=0.0179$, $J_3=0.0162$, and $J_4 =0.0118$.

When $I_{4}^{\left(\Delta_{-}, \Delta_{+}\right){\Re}}$ emerges as the dominant component among the $I_i$ terms (implying that $\mathcal{N}_{\Re\left(\mathcal{A}^*\mathcal{B}e^{i\theta}\right)}$ constitutes a remarkably large value among the $\mathcal{N}_\xi$ factors), the direct CP asymmetry $A_{CP}(\sqrt{s})$ undergoes a sign flip as the invariant mass $\sqrt{s}$ approaches $m_{\omega}$. In the standard phase-space integration process, this alternating behavior typically leads to severe numerical cancellations. To circumvent this difficulty, we introduce the $\operatorname{sgn}(x)$ sign function to transform the CP-violating observable from the conventional $A_{CP}$ into the modified $A_{CP}^{\Re}$. This approach effectively eliminates the potential cancellation effect, thereby successfully extracting the critical physics information embedded within the real part of the interference term.

In the decay process of $B^{-} \rightarrow \pi^{+} \pi^{-} \pi^{-}$, we obtained  $	I_{4}^{\left(\Delta_{-}, \Delta_{+}\right){\Re} }=3.556 \times 10^{-14}$, $I_{1}^{\left(\Delta_{-}, \Delta_{+}\right)}=3.444 \times 10^{-16}$, $I_{2}^{\left(\Delta_{-}, \Delta_{+}\right)}=-6.229 \times 10^{-15}$, $I_{3}^{\left(\Delta_{-}, \Delta_{+}\right)}=2.517 \times 10^{-14}$, which exhibits the properties of  $I_{4}^{\left(\Delta_{-}, \Delta_{+}\right){\Re}}$ is larger than $I_{1}^{\left(\Delta_{-}, \Delta_{+}\right)}$, $I_{2}^{\left(\Delta_{-}, \Delta_{+}\right)}$, $I_{3}^{\left(\Delta_{-}, \Delta_{+}\right)}$ and satisfies the  conditions for $I_{4}^{\left(\Delta_{-}, \Delta_{+}\right)}$ to be the dominant term. Since $I_{4}^{\left(\Delta_{-}, \Delta_{+}\right)}$ serves as a constituent component of the numerator of $A_{CP}$ while the corresponding denominator remains strictly positive, the overall mathematical characteristics of $A_{CP}$ are dictated by and approach those of $I_{4}^{\left(\Delta_{-}, \Delta_{+}\right)}$ when this term dominates.
It can be predicted that when the invariant mass of $\pi^+\pi^-$ in the decay process $B^{-} \rightarrow \pi^{+} \pi^{-} \pi^{-}$ is around $m_{\omega}$, the sign of $A_{CP}$ will change. Moreover, after introducing the $\operatorname{sgn}(x)$ sign factor, the integral value in the interference region is significantly higher than the result without this sign factor.

\section{The Amplitude of the Quasi-Two-Body Decay Process in Perturbative QCD}\label{Sec 3}
For the decay process $B^{-}\rightarrow\pi^{+}\pi^{-}\pi^{-}$, we compute the decay amplitudes $\mathcal{A}$ and $\mathcal{B}$ within the PQCD framework \cite{Keum:2000ms,pqcd2001-l,Keum:2000ph}, where $\mathcal{A}$ corresponds to the amplitude of $B^{-} \rightarrow \left(\rho^0 \rightarrow \pi^{+}\pi^{-}\right)\pi^{-}$, with the $\rho^0$ meson directly decaying into $\pi^{+}\pi^{-}$, and $\mathcal{B}$ corresponds to the amplitude of $B^{-} \rightarrow \left(\omega -\rho^{0} \rightarrow \pi^{+} \pi^{-} \right) \pi^{-}$, in which the $\pi^{+}\pi^{-}$ pair arises from the interference of $\omega$ and $\rho^0$ mesons, while higher-order contributions are neglected. 
 PQCD provides a theoretical formalism for analyzing strong interaction effects in the weak decays of $B$ mesons. The core concept involves factorizing the decay amplitude into three distinct components: a perturbatively calculable high-energy term known as the hard function $H$; a non-perturbative low-energy contribution determined by the meson wave function $\Phi$; and the Sudakov factor. The Sudakov factor enhances the convergence and reliability of the perturbative expansion by resumming leading logarithmic contributions, thereby mitigating endpoint divergences. This framework becomes especially effective when transverse momentum effects are incorporated 
\cite{Lu:2000em,Shi:2022ggo,Lu:2017ibo,Lu:2023yxa,Hua:2020usv,Yang:2022ebu,Li:2006jv}.

We determine the total amplitude by using the quasi-two-body decay method. To facilitate the analysis of interference effects, this amplitude does not contain the $s_{r}$ propagator.
The amplitudes $\mathcal{A}$ and $\mathcal{B}$ are specifically expressed in the PQCD framework as follows \cite{Yao:2024rgb}:
\allowdisplaybreaks
\begin{align}
	\mathcal{A} &= \sum_{\lambda=0, \pm 1} \frac{G_{F} P_{B^{-}} \cdot \epsilon^{*}(\lambda) g^{\rho^{0} \rightarrow \pi^{+} \pi^{-}} \epsilon(\lambda) \cdot\left(p_{\pi^{+}}-p_{\pi^{-}}\right)}{2} \nonumber \\
	&\quad \times \Bigg\{ V_{ub} V_{ud}^{*} \bigg\{ a_{1} \left[\mathcal{A}_{a b}^{L L}(\pi, \rho)+\mathcal{A}_{e f}^{L L}(\pi, \rho)-\mathcal{A}_{e f}^{L L}(\rho, \pi)\right] + a_{2} \mathcal{A}_{a b}^{L L}(\rho, \pi) 
 + C_{2}\left[\mathcal{A}_{c d}^{L L}(\pi, \rho)+\mathcal{A}_{g h}^{L L}(\pi, \rho)-\mathcal{A}_{g h}^{L L}(\rho, \pi)\right] \nonumber \\
 	&\quad + C_{1} \mathcal{A}_{c d}^{L L}(\rho, \pi) \bigg\} 
- V_{t b} V_{t d}^{*}\bigg\{ \left(a_{4}+a_{10}\right)\left[\mathcal{A}_{a b}^{L L}(\pi, \rho)+\mathcal{A}_{e f}^{L L}(\pi, \rho)-\mathcal{A}_{e f}^{L L}(\rho, \pi)\right] \nonumber \\
	&\quad + \left(a_{6}+a_{8}\right)\left[\mathcal{A}_{a b}^{S P}(\pi, \rho)+\mathcal{A}_{e f}^{S P}(\pi, \rho)-\mathcal{A}_{e f}^{S P}(\rho, \pi)\right] - \left(a_{4}-\frac{3}{2} a_{7}-\frac{3}{2} a_{9}-\frac{1}{2} a_{10}\right) \mathcal{A}_{a b}^{L L}(\rho, \pi) \nonumber \\
	&\quad + \left(C_{3}+C_{9}\right)\left[\mathcal{A}_{c d}^{L L}(\pi, \rho)+\mathcal{A}_{g h}^{L L}(\pi, \rho)-\mathcal{A}_{g h}^{L L}(\rho, \pi)\right]  + \left(C_{5}+C_{7}\right)\left[\mathcal{A}_{c d}^{S P}(\pi, \rho)+\mathcal{A}_{g h}^{S P}(\pi, \rho)-\mathcal{A}_{g h}^{S P}(\rho, \pi)\right] \nonumber \\
	&\quad - \left(C_{3}-\frac{3}{2} C_{10}-\frac{1}{2} C_{9}\right) \mathcal{A}_{c d}^{L L}(\rho, \pi) + \frac{3}{2} C_{8} \mathcal{A}_{c d}^{L R}(\rho, \pi) - \left(C_{5}-\frac{1}{2} C_{7}\right) \mathcal{A}_{c d}^{S P}(\rho, \pi) \bigg\} \Bigg\} , \\[15pt]
	\mathcal{B} &= \sum_{\lambda=0, \pm 1} \frac{G_{F} P_{B^{-}} \cdot \epsilon^{*}(\lambda) g^{\rho^{0} \rightarrow \pi^{+} \pi^{-}} \epsilon(\lambda) \cdot\left(p_{\pi^{+}}-p_{\pi^{-}}\right)}{2} \nonumber \\
	&\quad \times \Bigg\{ V_{ub} V_{ud}^{*} \bigg\{ a_{1} \left[\mathcal{A}_{a b}^{L L}(\pi, \omega)+\mathcal{A}_{e f}^{L L}(\pi, \omega)+\mathcal{A}_{e f}^{L L}(\omega, \pi)\right] + a_{2} \mathcal{A}_{a b}^{L L}(\omega, \pi) \nonumber \\
	&\quad + C_{2}\left[\mathcal{A}_{c d}^{L L}(\pi, \omega)+\mathcal{A}_{g h}^{L L}(\pi, \omega)+\mathcal{A}_{g h}^{L L}(\omega, \pi)\right] + C_{1} \mathcal{A}_{c d}^{L L}(\omega, \pi) \bigg\} \nonumber \\
	&\quad - V_{t b} V_{t d}^{*}\bigg\{ \left(a_{4}+a_{10}\right)\left[\mathcal{A}_{a b}^{L L}(\pi, \omega)+\mathcal{A}_{e f}^{L L}(\pi, \omega)+\mathcal{A}_{e f}^{L L}(\omega, \pi)\right] \nonumber \\
	&\quad + \left(a_{6}+a_{8}\right)\left[\mathcal{A}_{a b}^{S P}(\pi, \omega)+\mathcal{A}_{e f}^{S P}(\pi, \omega)+\mathcal{A}_{e f}^{S P}(\omega, \pi)\right] + \left(2 a_{3}+a_{4}+2 a_{5}+\frac{1}{2} a_{7}+\frac{1}{2} a_{9}-\frac{1}{2} a_{10}\right) \mathcal{A}_{a b}^{L L}(\omega, \pi) \nonumber \\
	&\quad + \left(C_{3}+C_{9}\right)\left[\mathcal{A}_{c d}^{L L}(\pi, \omega)+\mathcal{A}_{g h}^{L L}(\pi, \omega)+\mathcal{A}_{g h}^{L L}(\omega, \pi)\right]  + \left(C_{5}+C_{7}\right)\left[\mathcal{A}_{c d}^{S P}(\pi, \omega)+\mathcal{A}_{g h}^{S P}(\pi, \omega)+\mathcal{A}_{g h}^{S P}(\omega, \pi)\right] \nonumber \\
	&\quad + \left(C_{3}+2 C_{4}-\frac{1}{2} C_{9}+\frac{1}{2} C_{10}\right) \mathcal{A}_{c d}^{L L}(\omega, \pi) + \left(2 C_{6}+\frac{1}{2} C_{8}\right) \mathcal{A}_{c d}^{L R}(\omega, \pi) + \left(C_{5}-\frac{1}{2} C_{7}\right) \mathcal{A}_{c d}^{S P}(\omega, \pi) \bigg\} \Bigg\} .
\end{align}
 where  $g_{\rho^{0} \rightarrow \pi^{+} \pi^{-}}$ represents the coupling constant derived from the decay process of $\rho^{0} \rightarrow \pi^{+} \pi^{-}$ \cite{Bruch:2004py}. $V_{u b}, V_{u d}^{*}$ and $V_{t b}, V_{t d}^{*}$  are the fundamental elements of the CKM matrix. A momentum parameter $P_{i}$ (with $P_{i}$ representing $P_{{B}^{-}}$, $P_{\pi ^+}$, and $P_{\pi ^-}$) is defined, where $\epsilon$ stands for the polarization vector of the vector meson. The Fermi coupling constant is denoted by $G_F$, while $C_{i}$ represents the Wilson coefficient. Furthermore, the combinations $a_{i}$ are related to the Wilson coefficients $C_{i}$ \cite{Buchalla:1995vs}. $\text {LL, LR and SP refer to}\text { the contributions from }(V-A) \otimes(V-A),(V-A) \otimes(V+A) \text { and }(S-P) \otimes(S+P) \text { operators, respectively}$ \cite{Shi:2022ggo}. The symbol $\mathcal{A}_{ab}$ refers to the contribution originating from factorizable emission diagrams, and ${\cal A}_{cd}$ represents the contribution stemming from nonfactorizable emission diagrams. Similarly, ${\cal A}_{ef}$ (${\cal A}_{gh}$) denotes  the contribution of factorizable (nonfactorizable) annihilation diagrams. 
The parameters related to PQCD can be obtained from the literature \cite{Lu:2000em,Shi:2022ggo,Lu:2017ibo,Lu:2023yxa,Hua:2020usv,Yang:2022ebu,Li:2006jv,Yao:2024rgb}.

\section{Numerical Results and Analysis of CP Violation}\label{Sec 4}

The terms $V_{ub} V_{ud}^{*}$ and $V_{tb} V_{td}^{*}$ appearing in the equation are derived from the CKM matrix elements in the context of the Standard Model. These CKM matrix elements can be accurately determined through experimental measurements and are commonly parameterized using the Wolfenstein parameters $A$, $\rho$, $\lambda$ and $\eta$: $V_{ub} V_{ud}^{*}=A \lambda^{3}(\rho-i\eta)(1-\frac{\lambda^2}{2})$ and $V_{tb} V_{td}^{*}=A \lambda^{3}(1-\rho+i\eta)$ \cite{ParticleDataGroup:2024cfk}. The mixing parameter $\Pi_{\rho \omega}$, which has been precisely determined near the $\rho^0$ meson resonance, was recently reported by Wolfe and Maltman as  
$(-4470 \pm 250 \pm 160) - i(5800 \pm 2000 \pm 1100)~\mathrm{MeV}^{2}$ \cite{Wolfe:2009ts, Wolfe:2010gf}. The masses, decay widths, and other relevant parameters of the remaining particles are summarized in Table~\ref{TABLE:2}.

\begin{table}[h]
	\caption{Other parameters are from \ \cite{Yang:2022ebu,Cheng:2020ipp,ParticleDataGroup:2024cfk}}
	\label{tab:constant}
	\centering
	\begin{tabular*}{17cm}{@{\extracolsep{\fill}}lllll}
		\hline\hline\\
		\text{Mass(\text{GeV})}
		& $m_{B}=5.27934$     &$m_{\pi^{\pm}}=0.13957$   & $m_{\rho} = 0.77526$ & $m_{\omega} = 0.78265$            \\[3ex]
		\text{Wolfenstein parameters}
		& $\lambda =0.22650$  & $A=0.790$  &${\rho}=0.159$ & ${\eta}=0.352$ \\[3ex]
		\text{Decay constants (GeV)}
		&$f_{B} = 0.1900$ 
		& $f_{{\pi}} = 0.1302$ 
		& $f_{K} = 0.1557$   &$ f_{\phi}^{T} = 0.22$ \\[3ex]
		&$f_{\omega} = 0.195 $      &$f_{\phi} = 0.23 $      &$f_{\rho} = 0.209 $  &$f_{\omega}^{T} = 0.14 $ \\[3ex]
		\text{Decay width (\text{GeV})}
		& $\Gamma_{\rho} = 0.15 $ & $\Gamma_{\omega} = 8.49 \times 10^{-3}$        \\[3ex]
		\hline\hline
		\label{TABLE:2}
	\end{tabular*}
\end{table}

\begin{figure}[h]
	\centering
	\includegraphics[height=5.5cm,width=9cm]{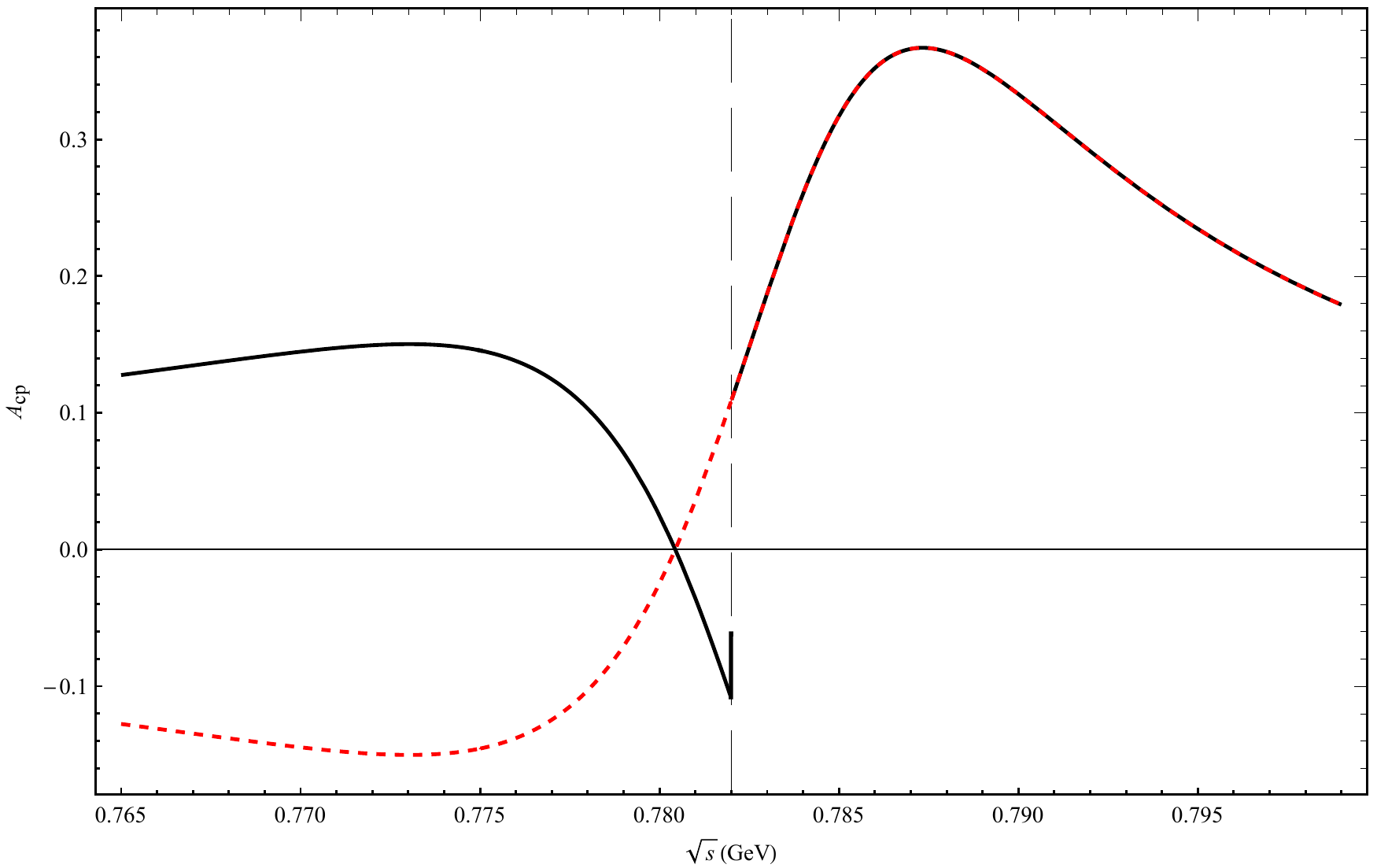}
	
	\caption{Plot of $ A_{C P} $  as a function of  $\sqrt{s}$  corresponding to central parameter values of CKM matrix elements and PQCD parameters.  The dotted line corresponds to the decay channel of $B^{-}  \rightarrow \pi^{+} \pi^{-} \pi^{-} \left(\operatorname{via} \: \rho^{0}-\omega\: \operatorname{mixing}\right)$. The solid line corresponds to the decay channel of $B^{-}  \rightarrow \pi^{+} \pi^{-} \pi^{-} \left(\operatorname{via} \: \rho^{0}-\omega\: \operatorname{mixing}\right)$ after adding $\operatorname{sgn}\left(x\right)$.}
	\label{fig:1}
\end{figure}

Based on the perturbative QCD amplitude expression derived in the previous study, together with the $\rho^{0}-\omega$ mixing mechanism and the corresponding input parameters, we present the variation of the asymmetry parameters $ A_{CP} $ and $ A_{CP}^{\Re} $ for the decay process $B^{-}  \rightarrow \pi^{+} \pi^{-} \pi^{-} \left(\operatorname{via} \: \rho^{0}-\omega\: \operatorname{mixing}\right)$ as a function of the invariant mass $\sqrt{s}$. The results are displayed in Fig.~\ref{fig:1}.

As shown by the dotted line in Fig.~\ref{fig:1}, consistent with our prediction in Sec. B of Sec.~\ref{Sec 2}, $ A_{CP} $ changes sign around $ \sqrt{s} = 0.782~\text{GeV} $, which can be directly attributed to the contribution of the $ I_{4}^{\left(\Delta_{-}, \Delta_{+}\right)} $ term. The CP violation central value in the decay channel $ B^{-} \rightarrow \pi^{+}\pi^{-}\pi^{-} $, calculated within the invariant mass region $ m_{\omega} \pm \Gamma_{\omega} $ using the central values of the CKM matrix elements and PQCD parameters, is $ 0.075 $. We argue that in the case of CP violation, both positive and negative deviations can serve as indicators of CP violation. In experimental settings, due to limitations in measurement precision and available data, CP violation effects in certain decay channels with large mass intervals may remain undetected as a result of such cancellations.

The solid black line segment $(A_{CP}^{\Re})$ represents the curve of the modified CP asymmetry that incorporates the $\operatorname{sgn}(s-m_{\omega}^{2})$ function. The vertical black dashed lines in the figure indicate the breakpoints of $A_{CP}^{\Re}$ caused by the $\operatorname{sgn}(x)$ sign function. By adding this piecewise function, the CP value is transformed from $A_{CP}$ to $A_{CP}^{\Re}$. The integral value of $A_{CP}^{\Re}$ within the interference region $m_{\omega} \pm \Gamma_{\omega}$ is $0.158$. This indicates that the numerical value of $A_{CP}^{\Re}$ has significantly increased compared to the value of $A_{CP}$ without the introduction of the $\operatorname{sgn}(x)$ sign function.

\begin{table}[h]
	{\renewcommand
		\scalebox{12}
		\centering %
		\renewcommand{\arraystretch}{2}
		\setlength{\tabcolsep}{5.5mm}{
			\begin{center}
				\caption{
					Different integral values of $ A_{CP} $, $A_{C P}^{\Re}$   obtained from various invariant mass ranges in the $B^{-}  \rightarrow \pi^{+} \pi^{-} \pi^{-} \left(\operatorname{via} \: \rho^{0}-\omega\: \operatorname{mixing}\right)$ decay process
				}
				\begin{tabular}{ ccc  }
					\hline
					Range of integration $\operatorname{\sqrt{s}}$$(\operatorname {GeV})$     & $A_{C P}$   &   $A_{C P}^{\Re}$   
					\\ \hline
					$[0.7735, 0.7800]$ & $-0.111_{-0.002-0.001}^{+0.003+0.002}$ &    $0.111_{-0.003-0.002}^{+0.002+0.001}$
					\\ \hline
					$[0.7825, 0.7905]$ & $0.301_{-0.012-0.016}^{+0.011+0.019}$ &    $0.301_{-0.012-0.016}^{+0.011+0.019}$
					\\ \hline
					$[0.7750, 0.7850]$ & $0.0187_{-0.004-0.002}^{+0.000+0.003}$ &    $0.101_{-0.003-0.002}^{+0.003+0.001}$
					\\ \hline	
					$[0.7735, 0.7905]$ & $0.0754_{-0.006-0.004}^{+0.000+0.005}$   &   $0.158_{-0.006-0.012}^{+0.004+0.011}$
					\\
					\hline
					\label{TABLE:1}
				\end{tabular}
	\end{center}}}
\end{table}

Within the PQCD framework, we present the numerical results of $A_{CP}$, 
$A_{C P}^{\Re}$ 
for different integration regions in Table~\ref{TABLE:1}. The first uncertainty originates from the CKM matrix, while the second one is associated with the PQCD method.
When $ m_{\omega} $ is defined as the sign-flip point, the $ A_{C P}^{\Re} $ within the interference region exhibits an integrated central value of $ 0.158 $, compared to the central value of $ 0.075 $ for $ A_{CP} $ over the invariant mass range from $ 0.7735 $ $\text{GeV}$ to $ 0.7905 $ $\text{GeV}$, where $ m_{\omega} = 0.782 $ $\text{GeV}$.

As presented in Table~\ref{TABLE:1}, a notable discrepancy arises between the integrated values computed with and without the $\operatorname{sgn}(x)$ function when the region of integration encompasses the zero-crossing point of $A_{CP}$. Within the resonance interference region, if the sign-function correction is omitted, the observed $A_{CP}$ value obtained through integration will be significantly suppressed whenever the selected integration interval straddles the sign-flip threshold—such as $\{0.7750\text{ GeV}, 0.7850\text{ GeV}\}$ and $\{0.7735\text{ GeV}, 0.7905\text{ GeV}\}$. This reduction is entirely attributed to the severe mutual cancellation between the positive and negative asymmetric contributions within the integration interval around the $\omega$ meson mass $m_\omega$.  Conversely, choosing asymmetric narrow kinematic windows that evade the zero-crossing point near $m_\omega$ (such as $\{0.7735\text{ GeV}, 0.7800\text{ GeV}\}$ and $\{0.7825\text{ GeV}, 0.7905\text{ GeV}\}$) yields a relatively large integrated CP asymmetry, but these intervals are far too restricted for practical experimental analyses. Under realistic experimental conditions, an insufficient sample size or limited statistics cannot satisfy such stringent narrow-window resolution requirements. Moreover, the presence of substantial background contamination from the broad $f_0(500)$ resonance further obscures the spectrum, rendering it exceptionally difficult to isolate a clear, distinct CP-violation signal driven by the narrow $\rho^0-\omega$ interference.  To completely resolve this dilemma and mathematically eliminate the cancellation effect without a loss of experimental statistics, we strongly recommend directly employing the physical mass threshold $m_{\omega}$ as the exact sign-flip point for the antisymmetric operator $\operatorname{sgn}(s - m_\omega^2)$. This choice is not only physically well-motivated but also maximizes the efficiency of extracting the localized CP-violation signal embedded within the real part of the $\rho^0-\omega$ interference amplitude.

\section{Summary and conclusion}\label{Sec 6}

In this work, a distinct methodological advantage of our proposed sign-function approach lies in its remarkable robustness against hadronic backgrounds originating from broad, smooth resonant states, particularly the $f_0(500)$ (or $\sigma$) meson. In standard experimental analyses of multi-body $B$ meson decays, such wide scalar resonances typically contribute a vast, slowly varying continuum background across the $\pi^+\pi^-$ invariant mass spectrum, severely obscuring the subtle quantum interference signals embedded within the narrow $\rho^0-\omega$ mixing region. Because the background amplitude from the $f_0(500)$ exhibits a flat and sign-preserving behavior over the localized $\rho^0-\omega$ interference window, the application of the antisymmetric  $\operatorname{sgn}(s-m_\omega^2)$ operator causes these smooth, non-interfering background terms to naturally cancel out upon phase-space integration. Consequently, this local algebraic method acts as an efficient ``hadronic filter,'' providing cleaner and more precise theoretical guidance for localized CP violation searches at future high-luminosity hadron colliders.

We have investigated CP violation arising from the real part of the interference term in the decay process $B^{-}  \rightarrow \pi^{+} \pi^{-} \pi^{-} \left(\operatorname{via} \: \rho^{0}-\omega\: \operatorname{mixing}\right)$  within the framework of Perturbative QCD. We find that within the same invariant mass range, the magnitude of the CP asymmetry is significantly enhanced when the $\operatorname{sgn}(x)$ sign factor is included compared to when it is not considered.  Meanwhile, we present the scope of applicability of the physical mechanism via $\rho^{0}-\omega$ mixing. 

Our analysis indicates that  $A_{CP}$ undergoes a sign flip when $I_{4}^{(\Delta_{-}, \Delta_{+})\Re}$ is relatively large compared to the other three terms and the invariant mass $\sqrt{s}$ approaches $m_{\omega}$. During the integration process, substantial cancellation may occur due to $\rho^{0}-\omega$ mixing. Therefore, decay processes that meet these conditions require the introduction of the $\operatorname{sgn}(x)$ function to convert $A_{CP}$ into $A_{CP}^{\Re}$, which allows for the detection of CP violation arising from $\rho^{0}-\omega$ mixing by eliminating such cancellation. Similar to Ref. \cite{zhangzhenhua2024}, the authors therein subtract the number of negative events  from the number of positive events of $A_{C P}^{F B}$ (CP caused by forward-backward asymmetry), and the resulting $A_{C P}^{F B, \mathfrak{I}}$ is significantly increased compared to the traditional method, which adds the number of positive examples to the number of negative examples. 
As a concrete example in experimental scenarios, consider a relatively narrow resonance interference region where $B^+$ decay events outnumber $B^-$ decay events in the first half of the invariant mass window, while $B^-$ decay events dominate in the second half, such that the net event excesses in both halves are equal in magnitude. If a single large bin is utilized to cover the entire interference region, the asymmetry between the total number of $B^+$ and $B^-$ events will be integrated out, causing the measured integrated CP asymmetry to inevitably vanish. This clearly explains why CP violation within such a narrow interference region could not be isolated under previous experimental statistics.
According to the prediction presented in this paper, future experiments with larger sample sizes may allow for further subdivision of the intervals, thereby enabling more effective observation of CP violation signals that could be obscured due to the mutual cancellation of positive and negative contributions at LHCb.

 \section*{Acknowledgements}
This work was supported by National Natural Science
Foundation of China under Grants Nos. 12475096 , Scientific Research Fund
of Hunan Provincial Education Department under Grants No. 22A0319.


\bibliographystyle{unsrt}  
\bibliography{refs}       

\end{spacing}
\end{document}